\documentclass[10pt,journal,cspaper,compsoc]{IEEEtran}
\ifCLASSOPTIONcompsoc
\else
  \usepackage{cite}
\fi
\ifCLASSINFOpdf
  \usepackage[pdftex]{graphicx}
  \graphicspath{{../pdf/}{../jpeg/}}
  \DeclareGraphicsExtensions{.pdf,.jpeg,.png}
  \graphicspath{ {./figure/} }
\else
\fi
\usepackage{array}

\usepackage{graphicx}
\usepackage{subfigure}
\begin{document}
%
\title{Real-time Adaptive Prediction Method for Smooth Haptic Rendering}
%
%
%
%

\author{Xiyuan~Hou,~\IEEEmembership{}
        Olga~Sourina,~\IEEEmembership{}
               
\IEEEcompsocitemizethanks{\IEEEcompsocthanksitem 

X. Hou is with the Fraunhofer IDM@NTU Center, Nanyang Technological University, NS1-1 Level 5, Nanyang Avenue 50, Singapore.
\protect\\
E-mail: houxy3@ntu.edu.sg
\IEEEcompsocthanksitem 
O. Sourina is with Fraunhofer IDM@NTU Center, Nanyang Technological University, NS1-1 Level 5, Nanyang Avenue 50, Singapore. \protect\\
E-mail: eosourina@ntu.edu.sg.
}
\thanks{}
}

%
%

\markboth{}%
{}
%


\IEEEcompsoctitleabstractindextext{%
\begin{abstract}
In this paper, we propose a real-time adaptive prediction method to calculate smooth and accurate haptic feedback in complex scenarios. Smooth haptic feedback is an important task for haptic rendering with complex virtual objects. However, commonly the update rate of the haptic rendering may drop down during multi-point contact in complex scenarios because high computational cost is required for collision detection and physically-based dynamic simulation. If the haptic rendering is done at a lower update rate, it may cause discontinuous or instable force/torque feedback. Therefore, to implement smooth and accurate haptic rendering, the update rate of force/torque calculation should be kept in a high and constant frequency. In the proposed method, the auto-regressive model with real-time coefficients update is proposed to predict interactive forces/torques during the physical simulation. In addition, we introduce a spline function to dynamically interpolate smooth forces/torques in haptic display according to the update rate of physical simulation. In the experiments, we show the feasibility of the proposed method and compare its performance with other methods and algorithms. The result shows that the proposed method can provide smooth and accurate haptic force feedback at a high update rate for complex scenarios. 
\end{abstract}

}

\maketitle

\IEEEdisplaynotcompsoctitleabstractindextext

%
\IEEEpeerreviewmaketitle

\section{Introductions}
%
%

%
%
%
%
\IEEEPARstart{I}{N} virtual reality (VR) technology, haptic interface is an active research field as the ``touch feeling'' plays an important role in manipulation of objects within a virtual environment. The real-time haptic rendering with complex or deformable objects is an important area of haptic research. During the haptic rendering, the forces applied to the haptic device must be updated at a high update rate (1 kHz) to maintain a realistic haptic simulation. The computational process of the collision detection with multiple contact points, deformable models, and the physically-based dynamic simulation is very time-consuming and the required high update rate of haptic rendering cannot be guaranteed at all time. Accurate and smooth force feedback is a challenging task for haptic rendering in complex virtual environments. Although the improvement of the computational ability of the hardware can speed up the haptic rendering, there is always a contradiction between high requirements of simulation and computing power of the computer. 

To realize high update rate haptic rendering in complex scenario, some multi-thread methods are proposed to separate the haptic thread with the physical thread \cite{no01,no02,no03}. These methods can improve the performance of the haptic rendering, but the stability and smoothness of the haptic force feedback is still not guaranteed \cite{no01}. When there are complex models, low update rate during collision detection and physical simulation may cause discontinuous forces applied to the user through the haptic device.

In this paper, we propose a real-time adaptive prediction method using interpolation to realize smooth and accurate haptic force feedback with complex models. An auto-regressive (AR) model is used to predict the future force from previous haptic force records. Meanwhile, the coefficients update algorithm updates the coefficients of the AR model in real-time to improve the accuracy of force prediction. In addition, a spline function is used to smoothly interpolate forces for haptic display in 1 kHz. The number of interpolated forces is calculated adaptively according to the speed of the physical simulation. This real-time prediction and interpolation process can provide smooth and accurate haptic feedback in complex virtual environments. 

The rest of the paper is organized as follows. The re-lated works are reviewed in Section 2. The real-time adaptive prediction method, interpolation function and the virtual coupling algorithm are described in Section 3.  The structure of the haptic rendering system is described in Section 4. Section 5 presents implementation of the proposed method and experiment results. The conclusion and future work are discussed in Section 6.

\section{Related Work}
There are some methods that have been developed to predict interactive haptic forces during haptic rendering. For linear prediction algorithms, Picinbono \cite{no04} proposed a linear extrapolation method to reconcile the update rate of the physically based deformable simulation and the haptic rendering. The linear extrapolation algorithms are used to predict the force in the haptic thread. In \cite{no05,no06,no07}, the real-time linear force extrapolation algorithm is implemented in the minimally invasive haptic surgery simulator to improve the stability of haptic rendering. As the deformable biomechanical model is simulated at a low rate of 30 Hz, the linear force prediction algorithm is used to achieve higher frequency update of haptic force feedback. Kim et al. \cite{no08} implemented a real-time haptic rendering system for deformable objects based on visual information (image obtained from a camera). They use the linear force extrapolation based on the position of the manipulator's tip to manage the different update rates of the visual thread and the haptic thread. Hu et al. \cite{no09} developed a magnetic haptic feedback system for surgical simulation and training.  A video-based tracking algorithm is implemented to get the position of the surgical-tool. The linear algorithm is used to extrapolate the position data to solve the problem of different update rates. In these proposed linear predictive algorithms, the inaccurate cases of overshoot and undershoot of the haptic force still exist during the haptic rendering.

In addition to the linear prediction algorithms, some AR models are applied to improve stability and synchronization of the haptic rendering system. In \cite{no10}, Wu et al. proposed a time series based prediction algorithm using the AR model to extrapolate forces for real-time haptic rendering with complex deformable model. The fixed AR coefficients are estimated from the user experience. During the haptic simulation, both the predicted force values and the simulated force values are used in the AR model. For physically-based medical simulation of the organs and tissues, Lee et al. \cite{no11} proposed to use a multi-rate estimator with time-varying parameters to improve the computational speed and accuracy. Although the accuracy of the haptic rendering is improved, the problems of discontinuous forces between successive haptic frames still exist. 

To realize smooth force feedback, Mazzella et al. \cite{no12} proposed a forcegrid data structure for the haptic force interpolation and extrapolation. In the forcegrid structure, the virtual workspace is divided into regular grids, and the force values are interpolated in each vertex. The haptic rendering algorithm can generate continuous forces in a higher update rate, regardless of the complexity of the models. But, this method only supports 3-DOF haptic rendering with one haptic device in a virtual environment. Fousek et al. \cite{no13} also presented a state-space haptic force pre-computation and approximation method based on the radial-basis function (RBF). The RBF interpolation could improve the accuracy of the approximation during the haptic interaction but the limitation is that they do not support dynamic virtual environment and deformable model. 

For stable haptic rendering, a virtual coupling model can obviously improve the stability by connecting the haptic device and the virtual tool with a spring link \cite{no14}. There are two types of virtual coupling models in haptic rendering systems: static virtual coupling and dynamic virtual coupling \cite{no15}. The static virtual coupling was introduced in \cite{no16,no17}. It is used for haptic rendering with virtual tools without physical property like mass. A quasi-static spring model is used to calculate stable haptic rendering. Although it can provide stable haptic rending in a high stiffness virtual environment, it only supports virtual tools without mass value. McNeely et al. \cite{no18} proposed the dynamic virtual coupling model in which user's motion and dynamic virtual tool were connected through a translational and rotational virtual springs. But, there are some computational instability problems in over stiff scenarios. In \cite{no19}, it was proposed a stable adaptive algorithm to provide stable and accurate haptic manipulation in dynamic virtual environment. The haptic rendering algorithm can automatically adjust virtual coupling parameters according to the mass values of the virtual tools. In addition, the force/torque magnitude can automatically saturate to the maximum force/torque values of the haptic device to provide stable haptic display.

\section{Methodologies}
The proposed method uses the real-time adaptive prediction model with coefficient update to calculate accurate interactive forces. Moreover, the interpolation algorithm can generate smooth force to haptic device in 1 KHz update rate. 

\begin{figure}[!t]
	\centering
	\includegraphics[width=0.48\textwidth]{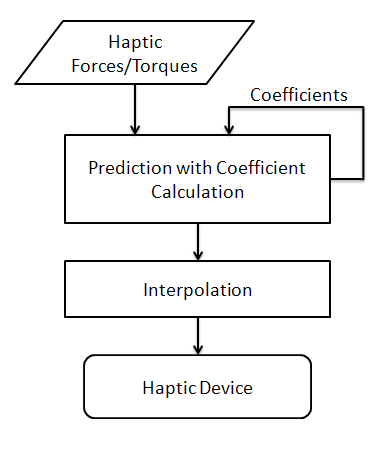}
	\caption{Structure of the real-time adaptive prediction method for haptic rendering. The coefficients of the prediction model are updated in real-time to generate accurate and smooth haptic force/torque feedback.}
	\label{fig:pipeline}
\end{figure}

For the AR model proposed in \cite{no20}, the coefficients need to be precomputed and fixed during the haptic rendering process. For different virtual tools and virtual objects, the AR model coefficients need to be changed to make sure accurate prediction of the haptic interaction forces. Otherwise, the accuracy and stability of haptic rendering may reduce. To overcome this drawback and improve the accuracy of the force prediction, in this paper we propose to update the coefficients of the AR model in real-time during the haptic rendering. Fig.~\ref{fig:pipeline} shows the structure of the real-time adaptive prediction method. At first, we use default AR coefficients for the haptic force prediction. It is similar with the prediction model with fixed coefficients proposed in \cite{no20}. When the haptic system recorded a certain number of force values, the algorithm calculates new coefficients based on the recorded interactive force value and updates the AR model. Therefore, the proposed method can predict more accurate force values during the process of haptic manipulation. 

In addition, the interpolation algorithm based on spline function is used to calculate smooth forces applied to the haptic device in the high update rate. The number of the interpolated force values is not fixed. It is calculated from the ratio of the high update rate of the haptic thread and the variable update of the physical simulation thread.

\subsection{Prediction with Coefficient Update}
\label{PredictionSection}
To realize accurate force prediction in real-time haptic rendering, we update the coefficients of the AR model to predict the haptic force in the next haptic frame. In the prediction calculation thread, on one hand, the coefficients calculation algorithm uses the recoded force values to calculation new coefficients. On the other hand, the AR model predicts the next force value through a linear combination of the current and previous force values with the updated coefficients. 

In the proposed haptic rendering method, $\mathbf{F_t}$ represents the current haptic force value calculated from the virtual coupling model. $\mathbf{F_t,F_{t-1},\ldots,F_{t-p+1}}$ represent previous force values which are stored during the haptic rendering process. Both current and previous forces values are used in the AR model to predict the next force value. The order $p$ of the AR model can be determined according to the Final Prediction Error (FPE) criterion. The next force  \({\bf{F}}_{t + 1|t}^p\) calculated from current time  with order of  is shown in the following equation:
\begin{equation}
{\bf{F}}_{t + 1|t}^p = {\phi _1}{\bf{F}}_t^{} + {\phi _2}{\bf{F}}_{t - 1}^{} +  \cdots  + {\phi _p}{\bf{F}}_{t - p + 1}^{} + {\varepsilon _{t + 1|t}}
\label{eq1}
\end{equation}

\[hjk = h1\sum {hj} \]

where \({\phi _1},{\phi _2}, \ldots ,{\phi _p}\) are new coefficients for haptic force prediction, \({\varepsilon _{t + 1|t}}\) is a white noise series with zero mean. 

For 6-DOF haptic rendering, the simulation needs to calculate and predict not only haptic force but also haptic torque. Therefore, torque prediction with AR model can be calculated using the following equation:
\begin{equation}
{\bf{T}}_{t + 1|t}^p = {\varphi _1}{\bf{T}}_t^{} + {\varphi _2}{\bf{T}}_{t - 1}^{} +  \cdots  + {\varphi _p}{\bf{T}}_{t - p + 1}^{} + {\varepsilon _{t + 1|t}}
\label{eq2}
\end{equation}
where \({\bf{T}}_{t + 1|t}^p\)  is the predicted next torque value,  are previous torque values. \({{\bf{T}}_t},{{\bf{T}}_t}, \ldots \;,{{\bf{T}}_{t - p + 1}}\) are new coefficients for haptic torque prediction. 

To improve the accuracy of the haptic force prediction, the coefficients calculated in real-time are used to update AR model during haptic rendering process. The least squares methods based on Yule-Walker equations is used to calculate the coefficients. At first, it is multiplied by  at both sides of (\ref{eq1})
\begin{equation}
{{\bf{F}}_{t - p + 1}}{{\bf{F}}_{t + 1|t}} = \mathop \Sigma \limits_{i = 1}^p \left( {{\phi _i}{{\bf{F}}_{t - p + 1}}{{\bf{F}}_{t - i + 1}}} \right) + {{\bf{F}}_{t - p + 1}}{\varepsilon _{t + 1|t}}.
\label{eq3}
\end{equation}
Then, we take the expectance of (\ref{eq3}) and eliminate the zero mean factors
\begin{equation}
E\left( {{{\bf{F}}_{t - p + 1}}{{\bf{F}}_{t + 1|t}}} \right) = \mathop \Sigma \limits_{i = 1}^p \left[ {{\phi _i}E\left( {{{\bf{F}}_{t - p + 1}}{{\bf{F}}_{t + 1|t}}} \right)} \right].
\label{eq4}
\end{equation}
Then the auto-covariance \(c_p\) and \(r_p\) auto-correlation   coefficients can be calculated as follows:
\begin{equation}
{c_p} = \frac{{E({{\bf{F}}_{t - p + 1}}{{\bf{F}}_{t + 1|t}})}}{{N - 1}} = \mathop \Sigma \limits_{i = 1}^p {\phi _i}{c_{i - p}}
\label{eq5}
\end{equation}
\begin{equation}
{r_p} = \frac{{{c_p}}}{{{c_0}}} = \mathop \Sigma \limits_{i = 1}^p {\phi _i}{r_{i - p}}
\label{eq6}
\end{equation}
Since that \({r_0} = 1\), the matrix form of the matrix of Yule-Walker equations \cite{no21} is described as follows:
\begin{equation}
\begin{array}{l}
\left[ {\begin{array}{*{20}{c}}
	1&{{r_1}}&{{r_2}}& \cdots &{{r_{p - 2}}}&{{r_{p - 1}}}\\
	{{r_1}}&1&{{r_1}}& \cdots &{{r_{p - 3}}}&{{r_{p - 2}}}\\
	\vdots & \vdots & \vdots &{}& \vdots & \vdots \\
	{{r_{p - 2}}}&{{r_{p - 3}}}&{{r_{p - 4}}}& \cdots &1&{{r_1}}\\
	{{r_{p - 1}}}&{{r_{p - 2}}}&{{r_{p - 3}}}& \cdots &{{r_1}}&1
	\end{array}} \right]\left[ {\begin{array}{*{20}{c}}
	{{\phi _1}}\\
	{{\phi _2}}\\
	\vdots \\
	{{\phi _{p - 1}}}\\
	{{\phi _p}}
	\end{array}} \right]\\
\quad \quad \quad \quad \quad \quad \quad  \quad \quad = \left[ {\begin{array}{*{20}{c}}
	{{r_1}}\\
	{{r_2}}\\
	\vdots \\
	{{r_{p - 1}}}\\
	{{r_p}}
	\end{array}} \right]
\end{array}
\label{eq7}
\end{equation}
where \(r_i\) is the autocorrelation coefficient at delay \(i\). The matrix and vector of \(r_i\) can be represented as \(\mathbf{R}\) and \(\mathbf{r}\). So, the (\ref{eq7}) also can be written as:
\begin{equation}
	{\bf{R}}{\kern 1pt} {\kern 1pt} {\bf{\Phi }} = {\bf{r}}
	\label{eq8}
\end{equation}
\begin{equation}
	{\bf{\Phi }} = {{\bf{R}}^{ - 1}}{\bf{r}}
	\label{eq9}
\end{equation}
In (\ref{eq8}) and (\ref{eq9}), \({\bf{\Phi }}\) is the vector of the auto-regressive co-efficients which can be solved with least-squares method.

\subsection{Smooth Interpolation }
\label{sec:3.2}
If the predicted haptic force is directly sent to the haptic device, the smoothness of the haptic display cannot be guaranteed. To generate continuous haptic display, the interpolation algorithm proposed in \cite{no20} is used to generate smooth force and torque feedback. Based on the predicted force values, the interpolation algorithm interpolates the haptic force values for haptic frames in 1 kHz. Meanwhile, these interpolated forces keep continuity with previous haptic force applied to the haptic device. 

To make sure continuity of the force display on haptic device, the interpolated forces are required to be curvature continuity with in successive haptic frames. The B-spline function is used to calculate the interpolated force values for haptic device because the B-spline function provides C2 continuity on the interpolated data and it doesn't pass the control points.  The interpolation function for the haptic force display is as follows:
\begin{equation}
	{\bf{F}}_{}^h(u) = {\bf{a}}{u^3} + {\bf{b}}{u^2} + {\bf{c}}u + {\bf{d}} 
\end{equation}
where $u$ is a parameter used to control the interpolated haptic forces applied to the haptic device. \({\bf{a}},{\bf{b}},{\bf{c}},{\bf{d}}\) are parameter vectors. We use the B-spline conditions for the interpolation. The function with parameters matrix is shown as follows:

\[{{\bf{F}}^h}(u) = \]
\begin{equation}
	\frac{1}{6}\left| {\begin{array}{*{20}{c}}
		{{u^3}}&{{u^2}}&{{u^{}}}&1
		\end{array}} \right|\left| {\begin{array}{*{20}{c}}
		{ - 1}&3&{ - 3}&1\\
		3&{ - 6}&3&0\\
		{ - 3}&0&3&0\\
		1&4&1&0
		\end{array}} \right|\left| {\begin{array}{*{20}{c}}
		{{\bf{F}}_i^p}\\
		{{\bf{F}}_{i + 1}^p}\\
		{{\bf{F}}_{i + 2}^p}\\
		{{\bf{F}}_{i + 3}^p}
		\end{array}} \right|
		\label{eq11}
\end{equation}
If the \({{\bf{M}}_s}\)  is the parameter matrix, \({{\bf{U}}^T}\) is the transpose of  vector. The haptic force and torque interpolation equations are:
\begin{equation}
	{{\bf{F}}^h}(u) = {{\bf{U}}^T}{{\bf{M}}_s}{{\bf{F}}^p},
	\label{12}
\end{equation}
\begin{equation}
	{{\bf{T}}^h}(u) = {{\bf{U}}^T}{{\bf{M}}_s}{{\bf{T}}^p}.
	\label{13}
\end{equation}
The value of  (from 0.0 to 1.0) is calculated as follows:
\begin{equation}
	u = i/n,\quad i \in [0,\;n - 1]
	\label{eq14}
\end{equation}
where $n$  represents the ratio of the update rate of the haptic thread and the update rate of the physical thread. During each period of physical simulation, $n$  is the number of the interpolated force values for haptic display. 

\begin{figure}[!t]
	\centering
	\includegraphics[width=0.48\textwidth]{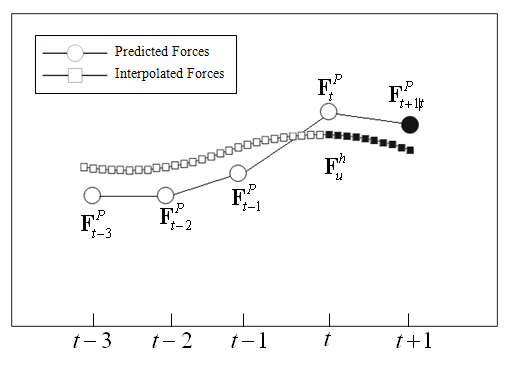}
	\caption{The force interpolation based on the predicted force values during the haptic rendering. The solid black circle \({\bf{F}}_{t + 1|t}^p\) represents the predicted next force at the current time $t$. }
	\label{fig:predict}
\end{figure}

The interpolation process based on the predicted force values is shown in Fig.~\ref{fig:predict}. The circle points rep-resent the predicted force values and the square points represent the interpolated force values. In the current haptic frame of time $t$, the next simulation force \({\bf{F}}_{t + 1|t}^P\) is calculated based on the prediction model. Then, the interpolation algorithm calculates the next force in the haptic frame from the predicted force values \({\bf{F}}_{t - 2}^p, \ldots ,{\bf{F}}_{t + 1|t}^P\). The \({\bf{F}}_u^h\) represents the interpolated force applied to the haptic device in the current haptic frame.

\subsection{Adaptive Virtual Coupling}
\begin{figure}[!t]
	\centering
	\includegraphics[width=0.48\textwidth]{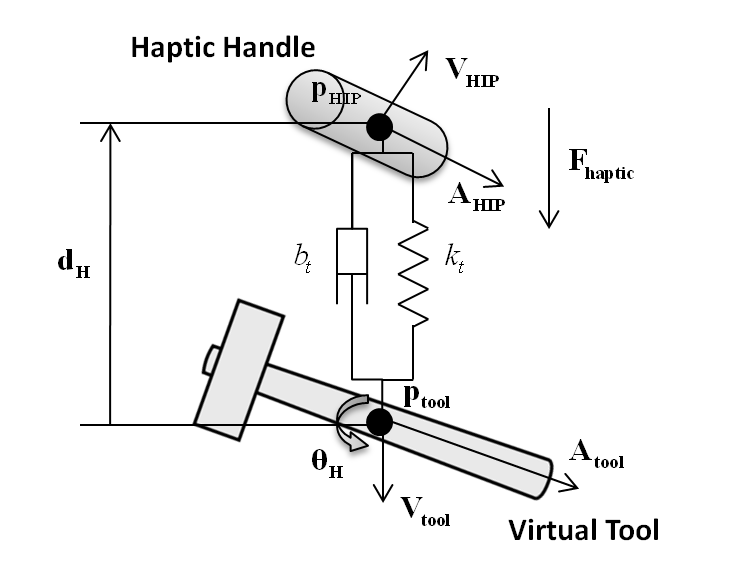}
	\caption{The adaptive virtual coupling model for 6-DOF haptic rendering \cite{no19}.}
	\label{fig:VC}
\end{figure}

For haptic rendering of the dynamic virtual tool with physical properties, we use the proposed stable adaptive haptic rendering algorithm based on virtual coupling to calculate the interaction forces and torques \cite{no19}. The virtual coupling model for 6-DOF haptic rendering is shown in Fig.~\ref{fig:VC}. The virtual tool and the haptic handle are connected through both translational spring and spiral spring models. The equation for interactive force calculation is as follows: 
\begin{equation}
{{\bf{F}}_{haptic}} =  - {k_t}({{\bf{P}}_{HIP}} - {{\bf{P}}_{tool}}) + {b_t}({{\bf{V}}_{HIP}} - {{\bf{V}}_{tool}})
\label{eq15}
\end{equation}
where \({{\bf{P}}_{HIP}}\) is the position of the haptic handle, \({{\bf{V}}_{HIP}}\) is the velocity of the haptic handle, \({{\bf{P}}_{tool}}\) is the position of the virtual tool and \({{\bf{V}}_{tool}}\) is the velocity of the virtual tool. The \(k_t\)  represents the stiffness parameter and  \(b_t\) represents the damping parameter. 

During the simulation, different virtual tools may have different mass values. To improve the accuracy and stability, the virtual coupling model should chose appropriate stiffness and damping parameters for each mass value $m$. In the stable adaptive algorithm \cite{no19}, the functions of \(k_t\) and \(b_t\) for each mass value is calculated as follows: 
\begin{equation}
	{k_t} = f(m),
	\label{eq16}
\end{equation}
\begin{equation}
	{b_t} = g({k_t},m).
	\label{eq17}
\end{equation}

For the torque calculation of the virtual coupling model, the interactive torque, the stiffness parameter $k_{rot}$  and the damping parameter $b_{rot}$ of the spiral spring is calculated from the similar equations.
\begin{equation}
\begin{array}{l}
{{\bf{T}}_{haptic}} =  - {k_{rot}}({{\bf{U}}_{HIP}} - {{\bf{U}}_{tool}})\\
\quad \quad \quad  \quad  + {b_{rot}}({{\bf{W}}_{HIP}} - {{\bf{W}}_{tool}})
\end{array},
\label{eq18}
\end{equation}
\begin{equation}
{k_{rot}} = d(I),
\label{eq19}
\end{equation}
\begin{equation}
{b_{rot}} = h({k_{rot}},I),
\label{eq20}
\end{equation}
where \({{\bf{U}}_{HIP}}\)  is the orientation of the haptic handle and  \({{\bf{U}}_{tool}}\) is the orientation of the virtual tool. \({{\bf{W}}_{HIP}}\)  is the angular velocity of the haptic device and \({{\bf{W}}_{HIP}}\) is the angular velocity of the virtual tool. $I$ is the inertia moment  of the virtual tool.

\section{Structure of Haptic Rendering}
\begin{figure}[!t]
	\centering
	\includegraphics[width=0.48\textwidth]{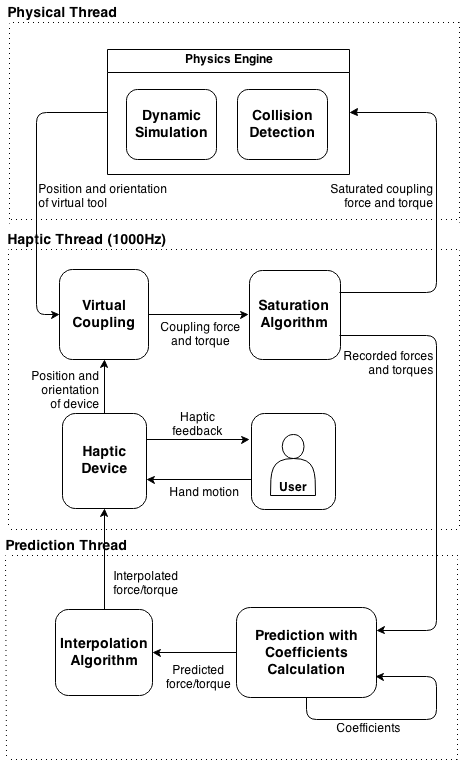}
	\caption{The pipeline of the haptic rendering system with physical thread, haptic thread and prediction thread. The haptic thread works at a constant 1 kHz. The update rate of the physical and prediction thread is changing during the haptic rendering.}
	\label{fig:renderpipeline}
\end{figure}

The pipeline of the proposed haptic rendering system is shown in Fig.~\ref{fig:renderpipeline}. It mainly includes three threads: physical thread, haptic thread, and prediction thread. These three threads work together to generate accurate and smooth haptic display.
The physical thread mainly calculates the collision detection, contact force and the corresponding dynamic movement of the virtual tool and other virtual objects. During the multi-point contact, the computation for collision detection and the dynamic simulation could be time consuming when the number of contact points is large. So, the update rate of the physical simulation calculated in the physical thread can be variable. Sometimes the update rate can be lower than 100 Hz.

In the haptic thread, first, position and orientation of the virtual tool and the configuration of the haptic device are used as inputs of the virtual coupling algorithm. The adaptive virtual coupling algorithm can adjust the stiffness and damping parameters for different mass values of virtual tools.  If virtual coupling forces are calculated at a lower rate about 100 Hz, the user will feel obvious lag and discontinuous force during haptic manipulation. So, there is a need for prediction and interpolation algorithms to calculate smooth haptic forces. 

In the prediction thread, the update rate is changing according to the update rate of the physical thread. The AR model is used to predict the next force value based on previous interaction forces. The coefficient calculation algorithm is implemented to update the coefficients of the AR model in real-time.  Meanwhile, the interpolation algorithm interpolates continuous force values using the B-spline functions. For each prediction calculation, the number of the interpolated force values is calculated from the ratio of the update rate of the haptic thread and physical thread.

\section{Experiment Results}

\subsection{Implementation}
In this section, the proposed real-time adaptive prediction haptic rendering method using interpolation is implemented in our haptic rendering system. Three benchmarks (peg-in-hole, Stanford bunny, and duck benchmarks) were implemented to analyze performance of the proposed haptic rendering method. 

The experiments are performed on a Windows PC with Intel Core2 Quad Q9400 2.66 GHz CPU. A PHANToM Premium 1.5/6DOF haptic device of SensAble Technologies is used in the experiment to provide both force/torque feedback. It also can be applied to other kinds of 3-DOF/6-DOF haptic device, such as Falcon haptic device of Novint Technologies. 

The haptic rendering system employs CHAI 3D API which provides both graphic and haptic interfaces. Different haptic devices can be used in the same virtual environment through the universal haptic interface of CHAI 3D library. It integrates Open Dynamic Engine (ODE) that provides an efficient computation of multi-point contact and physical simulation of multiple dynamic virtual objects \cite{no22}. Currently, available integrated haptic rendering algorithms of CHAI 3D library only provide 3-DOF haptic rendering. So, we implemented the proposed 6-DOF haptic rendering algorithm \cite{no23}.

\subsection{Accuracy Analysis}
\begin{figure}[!t]
		\centering
		\includegraphics[width=0.48\textwidth]{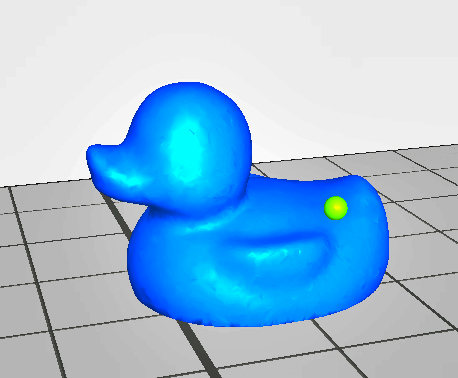}
		\caption{The duck benchmark (with 13926 polygons) used to evaluate the accuracy of the haptic rendering. }
		\label{fig:duck}
\end{figure}

In this experiment, the 3-DOF haptic rendering of the duck benchmark (shown in Fig.~\ref{fig:duck} is implemented to evaluate the accuracy of haptic rendering. We use the standard force data \cite{no24} to test the proposed prediction algorithms with/without coefficient update. 

The time complexity of the prediction algorithm with real-time coefficient calculation using the least-squares methods is $O({n^3})$. During the coefficient calculation, the size of the data window affects the computation time. Table~\ref{table:1} shows the experimental results of the data window size and the accuracy of prediction. When the size of the data window larger than 300 data samples, the improvement of the accuracy becomes very small. Therefore, we choose to use 300 values window size to recalculate coefficients of the prediction algorithm in real-time.
\begin{table}
\caption{Comparison of Accuracy with Different Window Size}
\begin{center}
	\begin{tabular}[c]{ c | c }
		\hline
		 Window Size & RMS Force Error (N) \\ \hline
		 100&	0.0563\\
		 200&	0.0388\\
		 300&	0.0306\\
		 400&	0.0306\\
		 500&	0.0302\\
		\hline
	\end{tabular}
	\label{table:1}
\end{center}
\end{table}

\begin{figure}[!t]
	\centering
	\includegraphics[width=0.48\textwidth]{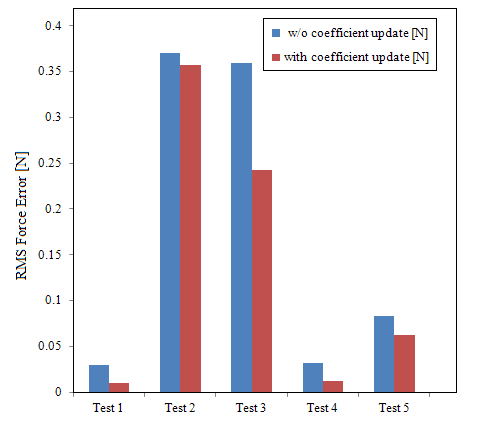}
	\caption{RMS force error of prediction algorithms with/without coefficients update.}
	\label{fig:forceError}
\end{figure}

To analyze the accuracy of the proposed real-time coefficient update algorithm, we select five force data sets from the duck benchmark for comparison. The one-way ANOVA test is performed on the force prediction results. Table~\ref{table:2} shows the comparison results of five tests. The small $p$ values indicate that the proposed prediction algorithm has significant improvement comparing with the previous prediction algorithm. Fig.~\ref{fig:forceError} shows the results of the RMS force error for prediction algorithms with/without coefficients update. From the experiment results, the RMS force error of the prediction algorithm with the coefficient update is less than the algorithm without coefficient update.
\begin{table}
	\caption{RMS Force Error with/without Coefficients Update}
	\begin{center}
		\begin{tabular}[c]{ c|c|c|m{0.08\textwidth}|m{0.08\textwidth} }
			\hline
			& F-value&	P-value&	RMS with w/o Coefficient Update [N]&  RMS with Coefficient Update [N]\\ \hline
			Test 1&	43.6&	$<$0.0001&	0.0291&	0.0099\\
			Test 2&	18.68&	$<$0.0001&	0.3707&	0.3569\\
			Test 3&	13.81&	0.0002&	0.3591&	0.2426\\
			Test 4&	7.64&	0.0057&	0.0321&	0.0125\\
			Test 5&	70.1&	$<$0.0001&	0.0835&	0.0627\\
			\hline
		\end{tabular}
		\label{table:2}
	\end{center}
\end{table}

\subsection{Stability and Smoothness Analysis}

In this section, we evaluate the stability and smoothness of the haptic feedback of the proposed real-time adaptive prediction method. Three different scenarios are used for haptic rendering evaluation. They are free space motion, general contact, and complex contact in a virtual environment scenario.

\subsubsection{Haptic Rendering in Free-Space}
\label{sec_5.3.1}

\begin{figure}[!t]
	\centering
	\includegraphics[width=0.48\textwidth]{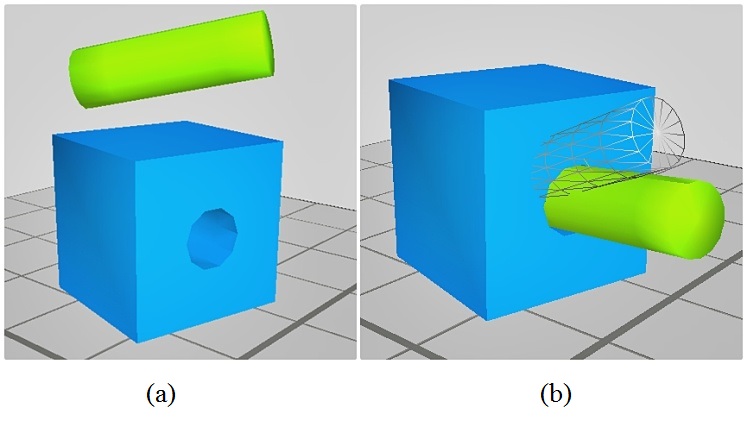}
	\caption{The peg-in-hole benchmark with 356 polygons. (a) The peg (used as the virtual tool) is in free-space. (b) The peg is inserted to the hole of the box. The wire-frame model shows the configuration of the real haptic device.}
	\label{fig:peg}
\end{figure}

\begin{figure}%
	\centering
	\subfigure[]{\includegraphics[width=0.48\textwidth]{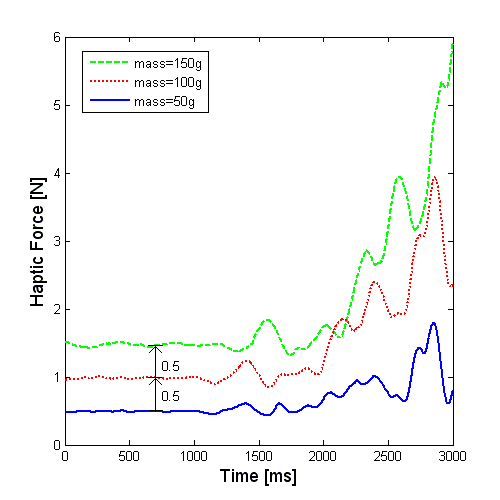} } 
	\subfigure[]{\includegraphics[width=0.48\textwidth]{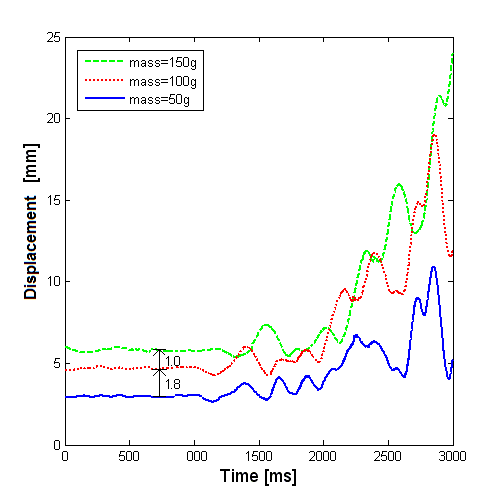} } 
	\caption{The haptic rendering of the peg-in-hole benchmark in the free-space. (a) The haptic force for the virtual tool with different mass values. (b) The displacement of the virtual tool during the haptic manipulation.}
	\label{fig8_ab}
\end{figure}

In this free space experiment, the peg-in-hole benchmark is implemented for evaluation, as shown in Fig.~\ref{fig:peg} The free-space means the virtual tool doesn't have any contact with virtual environment or other virtual objects \cite{no25}.

The purpose is to test the stability and force feedback of the proposed haptic rendering algorithm with different mass values in a free-space. The haptic device controls the peg that moves freely in the virtual environment without contact with other virtual objects. The box with the hole model is set to be static in the virtual environment. The proposed haptic rendering algorithm based on virtual coupling can adjust parameters of virtual coupling according to the mass value of the virtual tool.  The haptic device is connected with the peg through the virtual coupling model. The peg with three different mass values (50g, 100g, and 150g) is tested in the virtual environment without contact. 

Fig.~\ref{fig8_ab} shows the magnitude of haptic force and the displacement of the virtual tool during 3 seconds (3000 ms) movement of the haptic device in free space. The trajectory of the haptic device is per-recorded. In the experiment, for virtual tools with different mass values, the same trajectory is used to make comparisons.

In Fig.~\ref{fig8_ab}(a), from the 0 ms to 1000 ms, the haptic device is static in free space. Here for the bigger mass value of the virtual tool, the force applied to the haptic device is larger. From the 1000 ms to 3000 ms, the haptic device moves randomly in the free space. For the virtual tool with smaller mass value, the changes of the force magnitude are smaller than for bigger mass values as it is seen in the interval from 1500 ms to 3000 ms. For the virtual tool with larger mass value, the random movement generates larger changes of the force magnitude.

The Fig.~\ref{fig8_ab}(b) shows the displacement of the virtual tool with different mass values. When the mass values of virtual tools change (50 g, 100 g, and 150 g), the displacement (shown in Fig.~\ref{fig8_ab}(b)) does not increase linearly (shown in Fig.~\ref{fig8_ab}(a)). This is because the stable adaptive algorithm we use can adjust the stiffness parameter of the virtual coupling model automatically according to the physical properties of the virtual tool to improve the accuracy of the haptic manipulation \cite{no19}.

\subsubsection{Haptic Rendering during Contact}
\label{sec_5.3.2}
In this experiment, the peg-in-hole benchmark is used for evaluation of the haptic rendering during the contact. The purpose is to test the stability and smoothness of force feedback during a general contact scenario. The process is similar to the experiment in free space where different mass values of the virtual tool are used. In addition, the peg is inserted into the hole of the box as shown in Fig. 7(b). The trajectory of the virtual tool is per-recorded to make sure the same contact position is used for comparisons of the experiment of virtual tools with different mass values.

\begin{figure}[!t]
	\centering
	\subfigure[]{\includegraphics[width=0.49\textwidth]{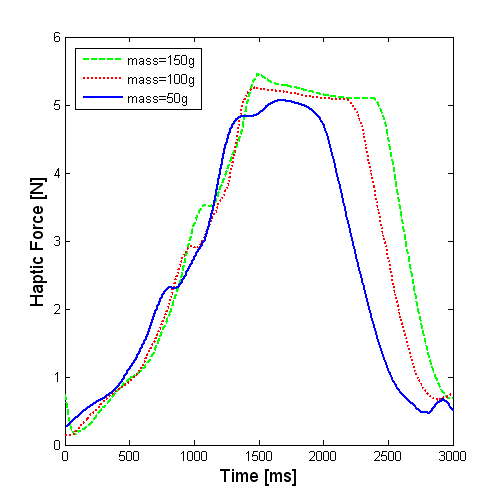} } 
	\subfigure[]{\includegraphics[width=0.49\textwidth]{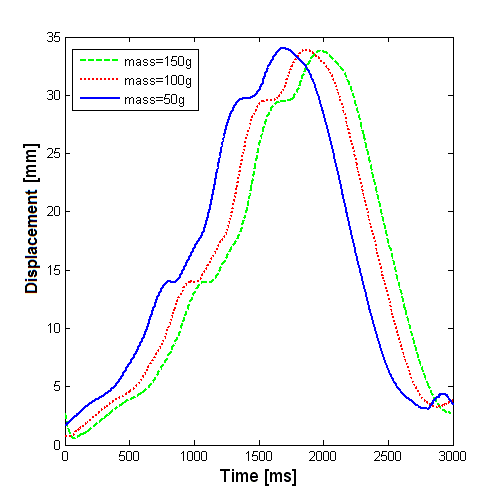} } 
	\caption{The haptic rendering of the peg-in-hole benchmark during the contact. (a) The haptic force for the virtual tools with different mass values. (b) The displacement of the virtual tool during the haptic manipulation.}
	\label{fig9_ab}
\end{figure}

Fig.~\ref{fig9_ab} shows the calculated haptic forces (as in Fig.~\ref{fig9_ab}(a)) and the corresponding displacement (as in Fig.~\ref{fig9_ab}(b)) of the virtual tool. For different mass values of virtual tools, the proposed haptic rendering algorithm can provide stable and smooth haptic feedback during the contact when the peg-in-hole benchmark is used. 

\subsubsection{Haptic Rendering in Complex Scenario}
\label{sec_5.3.3}

\begin{figure}[!t]
	\centering
	\includegraphics[width=0.48\textwidth]{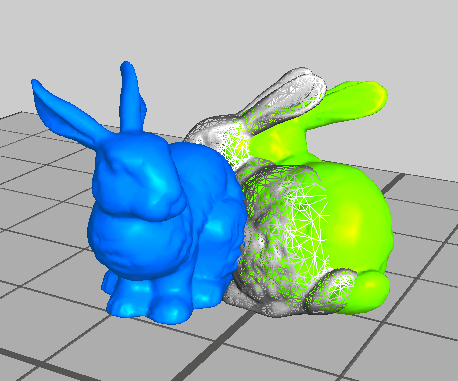}
	\caption{The Stanford bunny Benchmark implemented for haptic rendering in a complex scenario with multiple point contacts.}
	\label{fig_bunny}
\end{figure}

In the experiment with complex models, the Stanford bunny benchmark (shown in Fig.~\ref{fig_bunny}) is used to test the performance of the real-time adaptive prediction algorithm in a complex scenario with a large number of con-tact points. Another two haptic rendering algorithms (the stable adaptive algorithm without prediction \cite{no19} and the auto-regressive prediction algorithm without coefficient update \cite{no20}) are used for comparison with the proposed algorithm. During the experiment, one bunny is set to be static in the center of the virtual environment and another bunny is controlled by the haptic device to collide with the static bunny. In this scenario, each bunny is composed with 20989 polygons. 

The contact position is shown in Fig.~\ref{fig_bunny}. The blue bunny is static, and the green bunny represents the virtual tool. The current position of the haptic device is represented by a white wireframe of bunny model. During the experiment, the same trajectory is used to evaluate the algorithms. In this complex scenario, when collision happens, the update rate of physical simulation drops greatly.  For example, in our experiment, if the number of contact points increases to more than 20, the update rate of the physical simulation drops to 50 Hz. Comparing with 1 kHz update rate of haptic device, the low update rate of the physical simulation in the complex scenario causes discontinuous force feedback.

\begin{figure*}[!t]%
	\centering
	\subfigure[]{\includegraphics[width=0.48\textwidth]{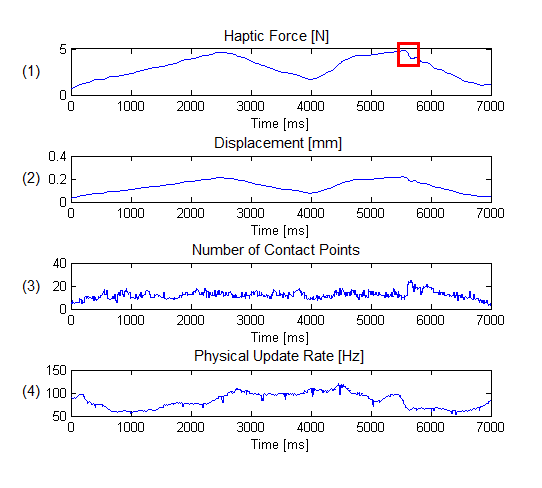} } 
	\subfigure[]{\includegraphics[width=0.48\textwidth]{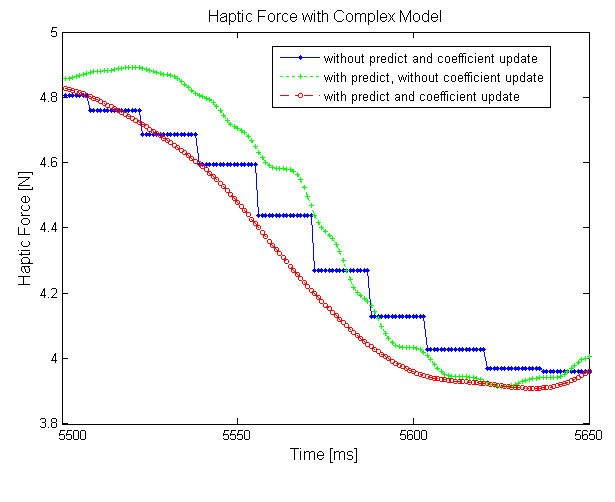} } 
	\caption{Haptic rendering with the Stanford bunny benchmark. (a) The performance of the haptic rendering during 7000 ms contact. It shows the haptic force, displacement of virtual tool, number of contact points and the physical thread update rate during contact. (b) The comparison of the smoothness of haptic rendering in the interval from 5500 to 5650.}
	\label{fig11_ab}
\end{figure*}

Fig.~\ref{fig11_ab}(a) shows performance of the proposed algorithm with the bunny benchmark. The haptic force magnitude, displacement, number of contact points, and the update rate of the physical simulation over time are shown in four rows. In Fig.~\ref{fig11_ab}(a), the number of the con-tact points (third row of Fig.~\ref{fig11_ab}(a)) is changing around 20 and the update rate of the physical thread (fourth row of Fig. 11(a)) is changing between 50 Hz to 150 Hz. In such low update rate, it is hard to maintain the force feedback stable and smooth as in the peg-in-hole benchmark. The proposed real-time adaptive prediction algorithm can maintain a general stable and smooth haptic display without obvious vibration and buzzing as seen in the first raw ofFig.~\ref{fig11_ab}(a).

To discuss details of the haptic forces, an interval from 5500 ms to 5650 ms is selected as seen in the first raw of Fig.~\ref{fig11_ab}(a) and the interval is presented in Fig.~\ref{fig11_ab}(b). It shows performances of three different haptic rendering methods that are used for haptic manipulation of the bunny model. In these haptic frames, the number of contact points is kept in the interval from 11 to 25, and the corresponding update rate of the physical simulation is changing in the interval 63 Hz to 72 Hz. For the haptic rendering method without prediction algorithm (in blue), the physical simulation results are used to calculate haptic force directly. There are obvious sharp changes on successive forces due to the low update rate of the physical simulation of the virtual tool.

The user could clearly feel the discontinuity of the force feedback when the haptic rendering method without prediction algorithm is used. For the prediction-based haptic rendering without coefficients update (in green), the force feedback is more smooth comparing with the haptic rendering method without force prediction. But, the stability of the force is still affected by the low update rate of the physical simulation. Finally, for the proposed haptic rendering method with prediction and real-time coefficients update (in red), the performance of the force feedback is more smooth and stable. Although the update rate of physical simulation is very low and keeps changing, the proposed method with adaptive prediction and interpolation improves the stability and smoothness of the haptic rendering.

\section{Conclusion}
\label{sec_6}
In a complex dynamic virtual environment, it is difficult to generate accurate and smooth haptic feedback when the physical simulation is done at a low update rate. In this paper, we proposed and developed a real-time coefficient update prediction algorithm to generate smooth haptic interaction force in high update rate. The auto-regressive model is used to predict the force value from the previous haptic force calculations. A real-time coefficient calculation algorithm is proposed to update the AR model during the haptic rendering. In addition, we implemented a spline function to interpolate force values for the haptic force feedback. The number of the interpolated forces is derived from the update rate of the physical simulation. The proposed algorithm is compared with other algorithms with standard benchmarks. It can overcome the force discontinuity caused by the heavy calculation of physical simulation or multiple point contacts with complex models.

\ifCLASSOPTIONcompsoc
  \section*{Acknowledgments}
\else
  \section*{Acknowledgment}
\fi

This research was done for Fraunhofer IDM@NTU, which is funded by the National Research Foundation (NRF) and managed through the multi-agency Interactive \& Digital Media Programme Office (IDMPO) hosted by the Media Development Authority of Singapore (MDA). This project is supported by the Ministry of Education of Singapore Grant MOE2011-T2-1-006 ``Collaborative Haptic Modeling for Orthopaedic Surgery Training in Cyberspace,''and by Russian Foundation for Basic Research 12-07-00678-a ``Research and Development of Force Interaction of Virtual Objects in the Tasks of Biomolecular Simulation.''

\ifCLASSOPTIONcaptionsoff
  \newpage
\fi



\bibliographystyle{IEEEtran}
\bibliography{hou_ref}
%



\end{document}